\documentclass[runningheads]{llncs}
\usepackage{graphicx}
\usepackage{amsmath}
\usepackage{amssymb}
\usepackage{multirow}
\usepackage{url}
\begin{document}
\title{Histogram Matching Augmentation for Domain Adaptation with Application to Multi-Centre, Multi-Vendor and Multi-Disease Cardiac Image Segmentation}
%
%
\author{Jun Ma\orcidID{0000-0002-9739-0855}}
\institute{Department of Mathematics, Nanjing University of Science and Technology\\ \email{junma@njust.edu.cn}}
\authorrunning{J. Ma, Histogram Matching Augmentation}
%
%
\maketitle              
\begin{abstract}
Convolutional Neural Networks (CNNs) have achieved high accuracy for cardiac structure segmentation if training cases and testing cases are from the same distribution. However, the performance would be degraded if the testing cases are from a distinct domain (e.g., new MRI scanners, clinical centers). In this paper, we propose a histogram matching (HM) data augmentation method to eliminate the domain gap. Specifically, our method generates new training cases by using HM to transfer the intensity distribution of testing cases to existing training cases. The proposed method is quite simple and can be used in a plug-and-play way in many segmentation tasks. The method is evaluated on MICCAI 2020 M\&Ms challenge, and achieves average Dice scores of 0.9051, 0.8405, and 0.8749, and Hausdorff Distances of 9.996, 12.49, and 12.68 for the left ventricular, myocardium, and right ventricular, respectively. 
Our results rank the third place in MICCAI 2020 M\&Ms challenge.
The code and trained models are publicly available at \url{https://github.com/JunMa11/HM_DataAug}.

\end{abstract}
\section{Introduction}
Accurate segmentation of the left ventricular cavity, myocardium and right ventricle from cardiac magnetic resonance images plays an import role for quantitative analysis of cardiac function, which can be used in clinical cardiology for patient management, disease diagnosis, risk evaluation, and therapy decision \cite{ESCGuidlines2019}.
In the recent years, many deep learning-based methods have achieved unprecedented performance (\cite{ACDC2018} \cite{chen2020CardiacReview}), especially when testing cases have the same distribution as training cases.
However, segmentation accuracy can be greatly degraded when these methods are tested on unseen datasets acquired from distinct MRI scanners or clinical centres \cite{RadiologyHeartMR}. This problem makes it difficult for these methods to be applied consistently across multiple clinical centres, especially when subjects are scanned using different MRI protocols or machines.

\begin{figure}
\center
\includegraphics[scale=0.4]{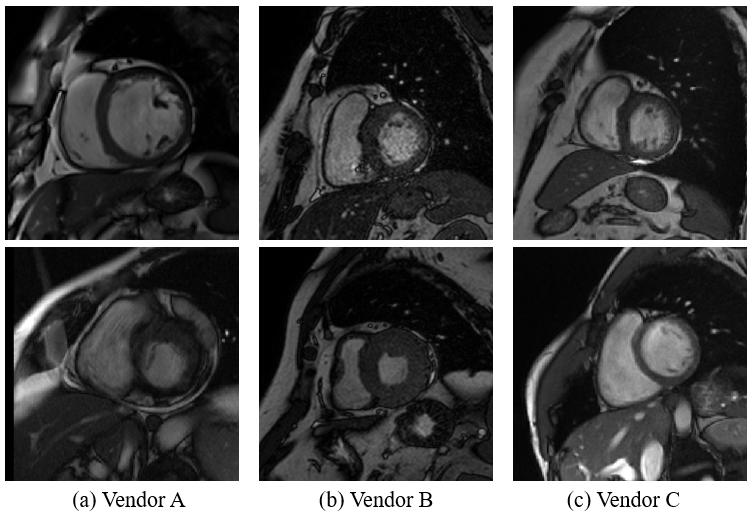}
\caption{Visual examples from different vendors. Images from different vendors have significant appearance variations.} \label{fig:examples}
\end{figure}

The M\&Ms challenge is the first international competition to date on cardiac image segmentation combining data from different centres, vendors, diseases and
countries at the same time. It evaluates the generalisation ability of machine/deep learning and cross-domain transfer learning techniques for cardiac image segmentation \cite{MMsZendo}. Figure \ref{fig:examples} shows six cardiac MR cases from three vendors. It can be observed that the appearance varies remarkably among different vendors.
Thus, how to develop a robust segmentation model that it can generalize to different centers, vendors, and diseases is an important but challenging problem.

Recently, many studies have been proposed to tackle this issue, such as domain adaptation (\cite{ganin2016NIPSdomAdv}, \cite{chen2019BrainDA}) and domain generalization \cite{douNIPS19domainGen}. Basically, domain adaptation aims to learn to align source and target domain in a domain-invariant high level feature space, which usually needs few annotated or unlabelled cases from the target domain during training. Domain generalization aims to train a model that it can directly generalize to new domains without need of retraining, which does not use data from the target domain.
In practice, both the two popular methods need to modify network architectures or loss functions.

Motivated by a recent study \cite{geirhos2018CNNbias} where CNNs are more sensitive to texture and intensity features. We aim to improve the generalization ability of CNNs by transferring the intensity distribution of the target dataset to the source dataset.
Specifically, we use histogram matching to bring the intensity appearance of the target dataset to the source dataset.
Instead of modifying the network architecture or loss function, our method only augments the training dataset, which is very simple and can be a plug-and-play method to any segmentation tasks.

\section{Proposed Method}
Histogram matching has been a widely used method, which generates a processed image with a specified histogram \cite{ImgPro}. We give a formal introduction of histogram matching as follows.
Let S and T denote continuous intensities (considered as random variables) of the source image and the target image, respectively. $P_S$ and $P_T$ denote their corresponding continuous probability density functions (PDF). We can estimate $P_S$ from the source image, and $P_T$ is the target probability density function. Let r be a random variable with the property
\begin{equation}\label{eq:S}
    r = M(S) = (L-1)\int_0^S P_S (x) dx,
\end{equation}
where $L$ is the number of intensity levels and x is a dummy variable of integration.
Suppose that a random variable $w$ has the property
\begin{equation}\label{eq:T}
    G(T) = (L-1)\int_0^T P_T (x) dx = r,
\end{equation}
it than follows from these two equations that $M(S) = G(T)$ and therefore, that $T$ must satisfy the condition
\begin{equation}\label{eq:S-T}
    T = G^{-1}[M(S)] = G^{-1}(r)
\end{equation}
Equations (\ref{eq:S})-(\ref{eq:S-T}) show that an image whose intensity levels have a specified probability density function can be obtained from a given image by using the following four steps:
\begin{itemize}
    \item Step 1. Obtain $P_S$ from the source image and use Eq. (\ref{eq:S}) to obtain the value of $r$.
    \item Step 2. Use the specified PDF in Eq. (\ref{eq:T}) to obtain the transformation function $G(T)$.
    \item Step 3. Obtain the inverse transformation $T = G^{-1}(r)$.
    \item Step 4. Obtain the output image by first equalizing the input image using Eq. (\ref{eq:S}); the pixel values in this image are the $r$ values. For each pixel with value $r$ in the equalized images, perform the inverse mapping $T=G^{-1}(r)$ to obtain the corresponding pixel in the output image. When all pixels have been thus processed, the PDF of the output image will be equal to the specified PDF.
\end{itemize}

\begin{figure}
\center
\includegraphics[scale=0.34]{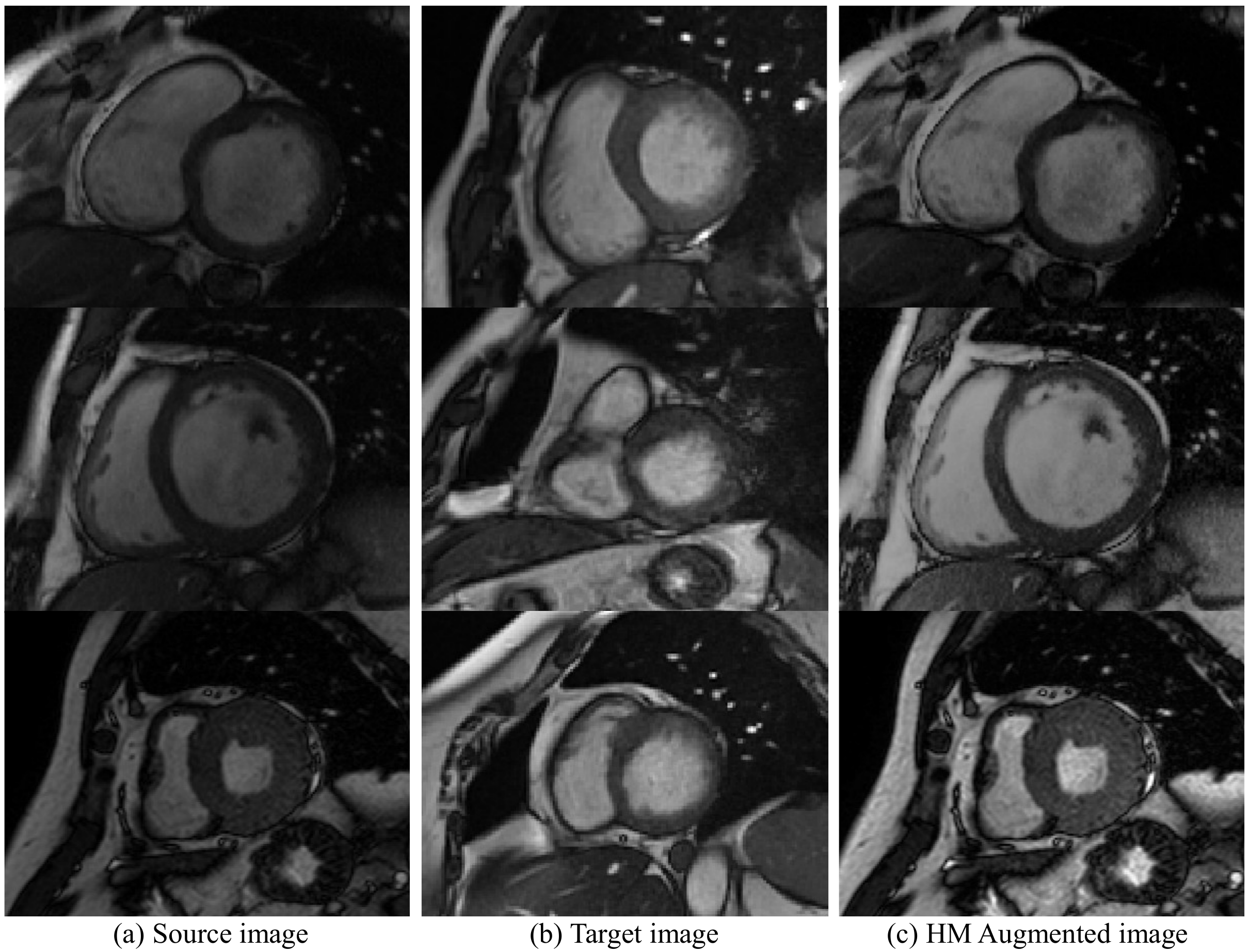}
\caption{Visual examples of the augmented images by histogram matching.} \label{fig:HM}
\end{figure}

Scikit-image \cite{skimage} has a build-in function \emph{match\_histograms}\footnote{https://scikit-image.org/docs/stable/} for histogram matching. In this paper, we use histogram matching to augment the training dataset so as to introduce the intensity distribution of the testing set. Specifically, we randomly select image pairs from labelled cases and unlabelled cases, and then transform the intensity distribution of the unlabelled case to labelled case. In this way, we can obtain many new training cases where its intensity distribution is similar to the unlabelled cases. Figure~\ref{fig:HM} presents some examples of the source images, target images and augmented images.

\section{Experiments and Results}
\subsection{Dataset and training protocols}
\subsubsection{Dataset}
The M\&Ms challenge cohort is composed of 350 patients with hypertrophic and dilated cardiomyopathies as well as healthy subjects. All subjects were scanned in clinical centres in three different countries (Spain, Germany and Canada) using four different magnetic resonance scanner vendors (Siemens, General Electric, Philips and Canon). The training set will contain 150 annotated images from two different MRI vendors (75 each) and 25 unlabelled images from a third vendor. The CMR images have been segmented by experienced clinicians from the respective institutions, including contours for the left (LV) and right ventricle (RV) blood pools, as well as for the left ventricular myocardium (MYO). The 200 test cases correspond to 50 new studies from each of the vendors provided in the training set and 50 additional studies from a fourth unseen vendor, that will be tested for model generalization ability. 20\% of these datasets will be used for validation and the rest will be reserved for testing and ranking participants. 


During preprocessing, we resample all the images to $1.25\times1.25\times8 mm^3$ and apply Z-score (mean subtraction and division by standard deviation)  to normalize each image. We employ nnU-Net \cite{nnunet20} as the default network. The patch size is $288\times288\times14$, and batch size is 8. We train 2D U-Net and 3D U-Net with five-fold cross validation. Each fold is trained on a TITAN V100 GPU with 1000 epochs. For each fold, we save the best-epoch model\footnote{Best-epoch model stands for the model that can achieves the best Dice on the validation set.} and final-epoch model. 
The code and trained models will be publicly available for research community after anonymous review. 
We declare that the segmentation method has not used any pre-trained models nor additional MRI datasets other than those provided by the organizers. 

\begin{figure}
\center
\includegraphics[scale=0.35]{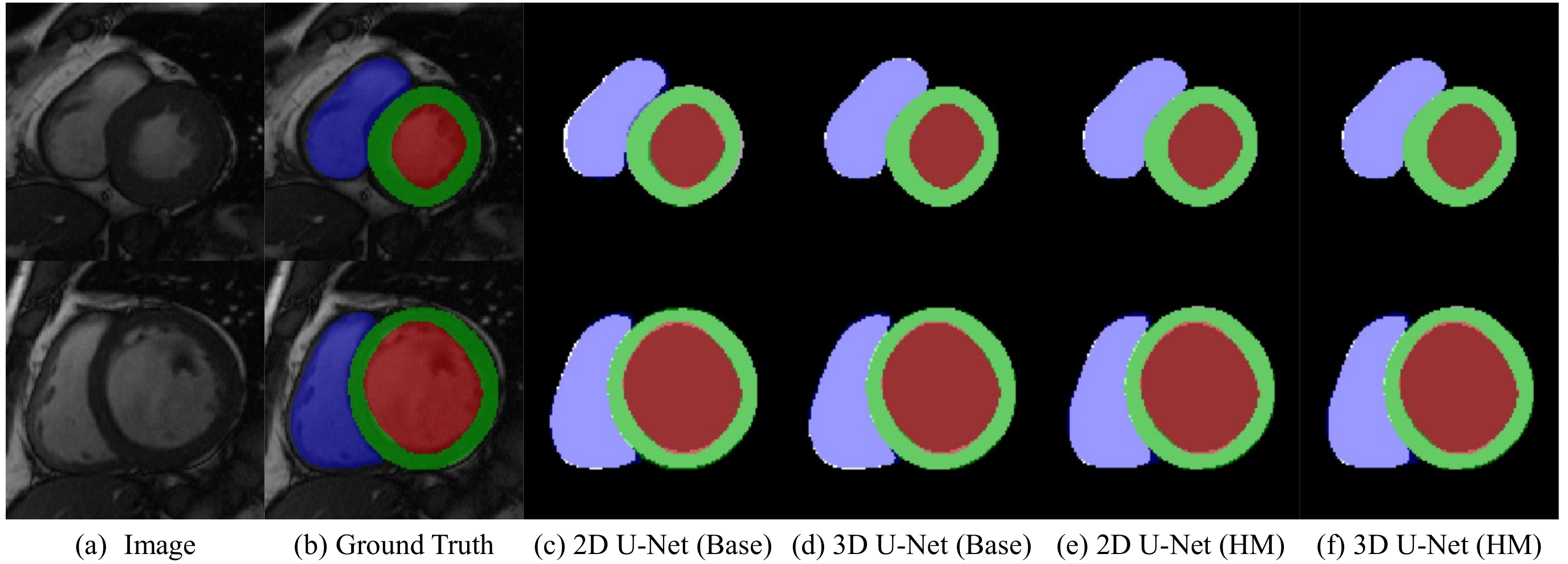}
\caption{Visual examples of segmentation results. Base and HM stand for the baseline dataset and the histogram matching augmented dataset.} \label{fig:SegRes}
\end{figure}

\subsection{Five-fold cross validation results}
Table~\ref{tb:5fold} presents the five-fold cross validation results of the final-epoch models of 2D U-Net and 3D U-Net\footnote{The results corresponding to the best-epoch model are not reported because it over-fits the validation set, where the corresponding Dice score is meaningless.}.
Basically, we found that the performance of 3D U-Net is slightly better than 2D U-Net. Figure~\ref{fig:SegRes} presents some visual segmentation results of different models. The methods achieve the best Dice scores for left ventricular, while the performance of myocardium is inferior than the left and right ventricular, indicating that the myocardium is more challenging to obtain accurate segmentation results.

\begin{table}[!htbp]
\caption{Five-fold cross validation results. It should be noted that The models trained on the default and the augmented dataset are not comparable, because the five-fold splits are different between the two datasets.}\label{tb:5fold}
\centering
\begin{tabular}{c|c|c|c|c|c}
\hline
Dataset                                                                                              & Model                     & Fold & LV\_Dice & Myo\_Dice & RV\_Dice \\ \hline
\multirow{10}{*}{\begin{tabular}[c]{@{}c@{}}Default\\ Dataset\\ (Baseline)\end{tabular}}             & \multirow{5}{*}{2D U-Net} & 0    & 0.9124   & 0.8695    & 0.8842   \\
                                                                                                     &                           & 1    & 0.9254   & 0.8756    & 0.9074   \\
                                                                                                     &                           & 2    & 0.9218   & 0.8753    & 0.8945   \\
                                                                                                     &                           & 3    & 0.9265   & 0.8684    & 0.8845   \\
                                                                                                     &                           & 4    & 0.9291   & 0.8645    & 0.8874   \\ \cline{2-6} 
                                                                                                     & \multirow{5}{*}{3D U-Net} & 0    & 0.9256   & 0.8873    & 0.8981   \\
                                                                                                     &                           & 1    & 0.9344   & 0.8879    & 0.9141   \\
                                                                                                     &                           & 2    & 0.9399   & 0.8828    & 0.9010   \\
                                                                                                     &                           & 3    & 0.9372   & 0.8760    & 0.8930   \\
                                                                                                     &                           & 4    & 0.9308   & 0.8753    & 0.8944   \\ \hline
\multirow{10}{*}{\begin{tabular}[c]{@{}c@{}}Histogram\\ Matching\\ Augmented\\ Dataset\end{tabular}} & \multirow{5}{*}{2D U-Net} & 0    & 0.9832   & 0.9639    & 0.9762   \\
                                                                                                     &                           & 1    & 0.9871   & 0.9729    & 0.9825   \\
                                                                                                     &                           & 2    & 0.9871   & 0.9747    & 0.9803   \\
                                                                                                     &                           & 3    & 0.9831   & 0.9683    & 0.9784   \\
                                                                                                     &                           & 4    & 0.9835   & 0.9735    & 0.9772   \\ \cline{2-6} 
                                                                                                     & \multirow{5}{*}{3D U-Net} & 0    & 0.9895   & 0.9633    & 0.9760   \\
                                                                                                     &                           & 1    & 0.9887   & 0.9690    & 0.9826   \\
                                                                                                     &                           & 2    & 0.9824   & 0.9632    & 0.9719   \\
                                                                                                     &                           & 3    & 0.9871   & 0.9654    & 0.9796   \\
                                                                                                     &                           & 4    & 0.9870   & 0.9741    & 0.9762   \\ \hline
\end{tabular}
\end{table}

\subsection{Validation set results}
The validation set is hidden by challenge organizers. We package our code and model in a singularity container and submit it to the organizers. Due to the limited number of submission tries, we do not submit the baseline models because the models trained on augmented datasets would be better than baseline models.
In summary, we submit the following five solutions on the validation set.
\begin{itemize}
    \item Solution 1. 3D U-Net best-epoch model;
    \item Solution 2. 3D U-Net final-epoch model;
    \item Solution 3. Ensemble of 3D U-Net best-epoch model and 2D U-Net best-epoch model;
    \item Solution 4. Ensemble of 3D U-Net final-epoch model and 2D U-Net final-epoch model;
    \item Solution 5. Ensemble of the above four solutions;
\end{itemize}

Table~\ref{tb:val} shows the quantitative results of the five solutions on validation set. It can be observed that assembling multiple models may improve Dice, but would degrade the HD and ASSD. We also apply paired T-test between the solution 1 and the other four solutions to show whether their performances are statistically significant different. Surprisingly, the statistical significance level $p>0.05$  for all comparisons. In other words, the performances of solution 2-4 do not have statistically significant difference compared with the solution 1.
It can be found that ensemble more models can obtain sightly better Dice scores, but could degrade the Hausdorff distance.
We also compare our results with the recent work \cite{li2020GAN-MMs} that uses GAN for domain adaptation, and our methods obtain better Dice scores for LV, Myo and RV.
Finally, we select the solution 1 as our final solution for the hidden testing set.

\begin{table}[!h]
\caption{Quantitative results of different solutions on the official hidden validation set. `-' denotes not reported.}\label{tb:val}
\centering
\begin{tabular}{l|ccc|ccc|ccc}
\hline
\multirow{2}{*}{Solution} & \multicolumn{3}{c|}{LV} & \multicolumn{3}{c|}{Myo} & \multicolumn{3}{c}{RV}   \\ \cline{2-10} 
                           & Dice   & HD    & ASSD  & Dice    & HD     & ASSD  & Dice   & HD     & ASSD   \\ \hline
Solution 1                & 0.9130  & \textbf{8.089} & 1.017 & 0.8627  & \textbf{12.19} & \textbf{0.710} & \textbf{0.8937} & 11.88 & \textbf{0.9078} \\
Solution 2                & 0.9131  & 8.091 & \textbf{1.015} & 0.8627  & 12.20 & \textbf{0.710} & 0.8935 & 11.85 & 0.9101 \\
Solution 3                & \textbf{0.9166}  & 8.090 & 0.958 & \textbf{0.8660}  & 12.25 & 0.734 & 0.8927 & 11.35 & 0.9682 \\
Solution 4                & \textbf{0.9166}  & 8.103 & 0.959 & 0.8658  & 12.26 & 0.736 & 0.8927 & \textbf{11.34} & 0.9667 \\
Solution 5                & \textbf{0.9166}  & 8.099 & 0.959 & 0.8659  & 12.25 & 0.735 & 0.8927 & 11.35 & 0.9674 \\ \hline
GAN \cite{li2020GAN-MMs} & 0.903 &-&-&0.859 & -&- & 0.865&-&- \\
\hline
\end{tabular}
\end{table}

\subsection{Testing set results}
Table \ref{tb:test} presents the average Dice, HD and ASSD for each vendor on testing set.
Overall, the performances on vendor C and D are lower than the performances on vendor A and B, because the training set does not have annotated cases from the vendor C and D.
It can be observed that the performance on the vendor C is better the the performance on the vendor D, especially in HD and ASSD with improvements up to $5mm$, which could\footnote{Here we use `could' because we do not have corresponding testing results of our baseline model where histogram matching data augmentation is not used.} demonstrate the effectiveness of our histogram matching data augmentation.

\begin{figure}
\centering
\includegraphics[scale=0.35]{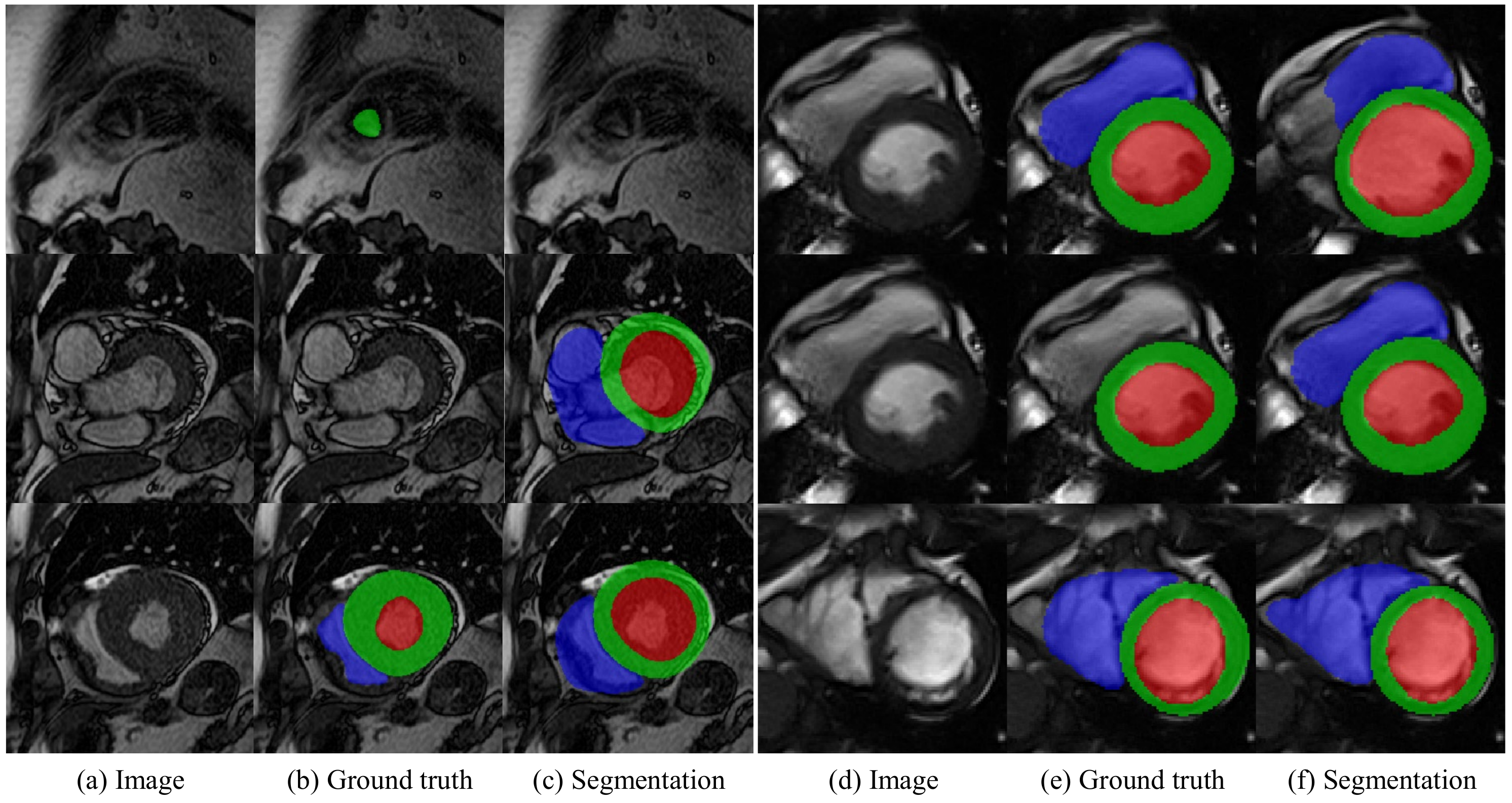}
\caption{Visual examples of worse cases in vendor A and B.} \label{fig:worse-AB}
\end{figure}

\begin{table}[!h]
\caption{Quantitative results on the official hidden testing set}\label{tb:test}
\centering
\begin{tabular}{c|ccc|ccc|ccc}
\hline
\multirow{2}{*}{Vendor} & \multicolumn{3}{c|}{LV} & \multicolumn{3}{c|}{Myo} & \multicolumn{3}{c}{RV}  \\ \cline{2-10} 
                        & Dice   & HD     & ASSD  & Dice    & HD     & ASSD  & Dice   & HD     & ASSD  \\ \hline
A                       & 0.9148 & 10.78 & 1.003 & 0.8435  & 14.14 & 0.697 & 0.8771 & 12.84 & 1.142 \\
B                       & 0.9136 & 7.866  & 0.971 & 0.8675  & 10.14 & 0.717 & 0.8792 & 11.61 & 1.144 \\
C                       & 0.8943 & 9.231  & 1.389 & 0.8265  & 11.33 & 1.066 & 0.8732 & 10.82 & 1.152 \\
D                       & 0.8977 & 12.11 & 1.521 & 0.8243  & 14.34 & 0.958 & 0.8703 & 15.46 & 1.513 \\ \hline
All                     & 0.9051 & 9.996  & 1.221 & 0.8405  & 12.49 & 0.859 & 0.8749 & 12.68 & 1.238 \\ \hline
\end{tabular}
\end{table}

\subsubsection{Worse cases analysis}
\begin{figure}
\centering
\includegraphics[scale=0.35]{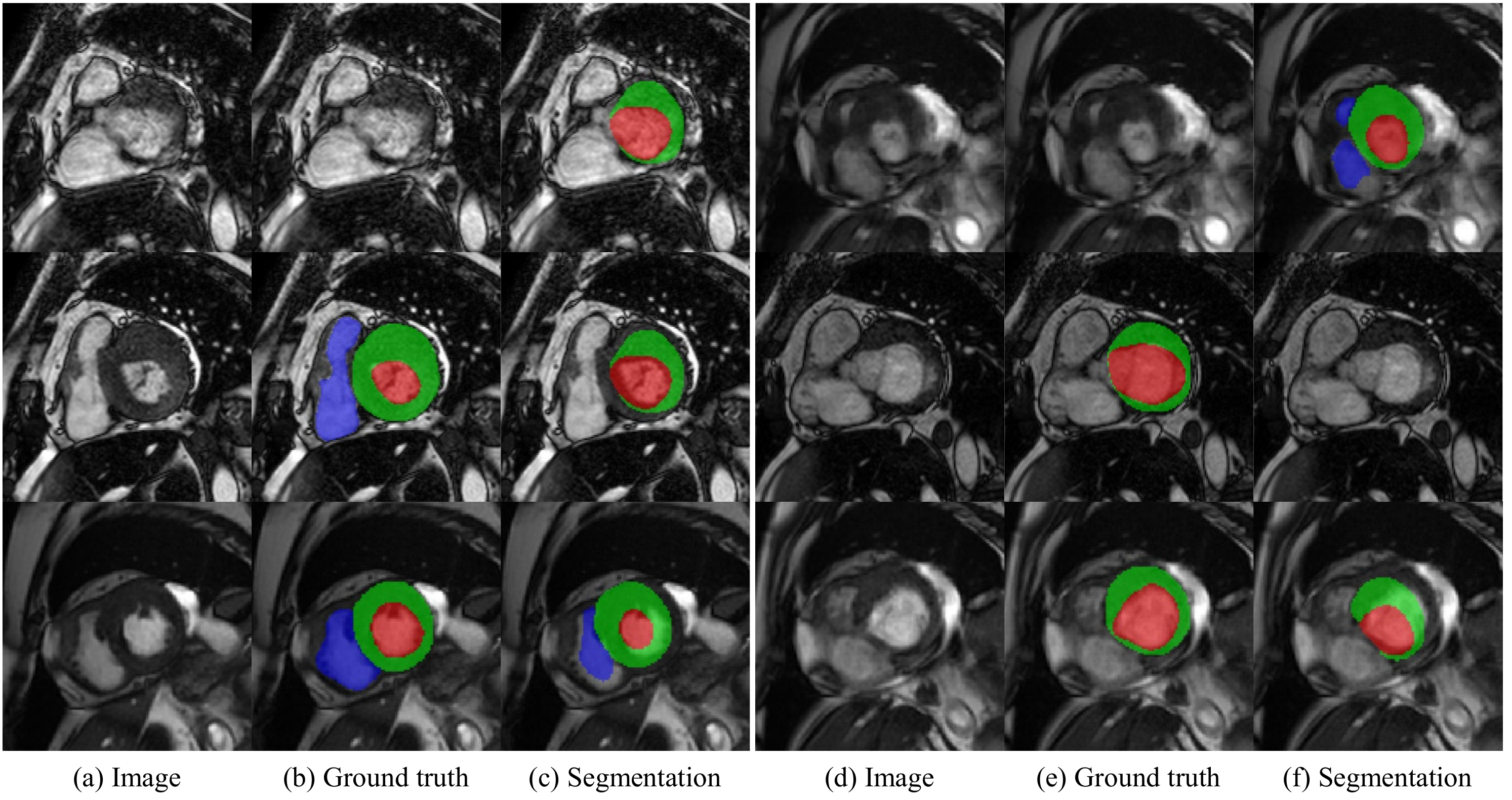}
\caption{Visual examples of worse cases in vendor C and D.} \label{fig:worse-CD}
\end{figure}

Figure~\ref{fig:worse-AB} and~\ref{fig:worse-CD} show some segmentation results with low performance.
Basically, it can be observed that most of the segmentation errors occur in the top or bottom of the heart because these regions usually have low contrast and ambiguous boundaries.
We find that the worse segmentation results can also have reasonable shapes even for the severe over-segmentation (e.g., Figure~\ref{fig:worse-AB} the 2nd row ).
Some LV segmentation results are significantly smaller or larger than ground truth, which could motivate us to improve our network by imposing size constrain, such as constraining the volume of network outputs to be within a specified range.

\subsubsection{Limitation}
Although histogram matching is a good pre-processing technique to eliminate intensity distribution differences, it should be noted that histogram matching could cover specific characteristics of other MRI modalities, like for example, LGE MRI. With the histogram changes, relevant informations about scar tissue might be degraded. In the future, we need to verify the effects of histogram matching data augmentation on multi-sequence cardiac MR datasets, such as MyoPS \cite{MyoPS}.

\section{Conclusion}
One of the challenging problems of current segmentation CNNs is that the performance would degrade when applying the trained model to a new dataset.
In this paper, we introduce histogram matching to augment the training cases that have the similar intensity distributions to the new (unlabelled) dataset set, which is very simple and can be a plug-and-play method to any segmentation CNNs.
Based on the quantitative results on the validation set, we believe that our method can be a strong baseline.

\section*{Acknowledgement}
The authors of this paper declare that the segmentation method they implemented for participation in the M\&Ms challenge has not used any pre-trained models nor additional MRI datasets other than those provided by the organizers.
We also thank the organizers for hosting the great challenge.

%
%
%
\bibliographystyle{splncs04}
\bibliography{Ref}

\end{document}